\documentclass[conference]{IEEEtran}
\usepackage{lipsum}
\usepackage{balance}
\usepackage[utf8]{inputenc} %
\usepackage{hyperref}       %
\usepackage{url}            %
\usepackage{booktabs}       %
\usepackage{amsfonts}       %
\usepackage{nicefrac}       %
\usepackage{microtype}      %
\usepackage{xcolor}
\usepackage{amsmath}
\usepackage{amsthm}
\usepackage{amssymb}
\usepackage{graphicx}
\usepackage{tikz}
\usepackage[labelfont=bf]{caption}
\usepackage{multirow}
\usepackage{wrapfig}
\usepackage{enumitem}
\usepackage{subcaption}
\usepackage{soul}
\usepackage{bm}
\usepackage{wrapfig}
\usepackage{listings}
\usepackage{todonotes}
\usepackage{jlcode}
\usepackage[linesnumbered,ruled,vlined]{algorithm2e}
\usepackage[bottom]{footmisc}
\usepackage[makeroom]{cancel}

\definecolor{mydarkblue}{rgb}{0,0.08,0.45}
\definecolor{myfavblue}{rgb}{0.1176, 0.392, 1.0}
\hypersetup{
    colorlinks=true,
    linkcolor=mydarkblue,
    citecolor=green,
    filecolor=mydarkblue,
    urlcolor=mydarkblue}

\def\BibTeX{{\rm B\kern-.05em{\sc i\kern-.025em b}\kern-.08em
    T\kern-.1667em\lower.7ex\hbox{E}\kern-.125emX}}

\begin{document}

\title{RayTracer.jl: A Differentiable Renderer that supports Parameter Optimization for Scene Reconstruction}

\author{
    Avik Pal\\
    Computer Science and Engineering\\
    Indian Institute of Technology Kanpur\\
    avik.pal.2017@gmail.com
}

\maketitle

\begin{abstract}
In this paper, we present RayTracer.jl, a renderer in Julia that is fully differentiable using source-to-source Automatic Differentiation (AD). This means that RayTracer not only renders 2D images from 3D scene parameters, but it can be used to optimize for model parameters that generate a target image in a Differentiable Programming (DP) pipeline. Our renderer is a complete general purpose renderer, which means that unlike most previous work, we do not make any changes to the renderer to make it differentiable. Additionally, we interface our renderer with the deep learning framework Flux for use in combination with neural networks. In this paper, we also demonstrate the use of this differentiable renderer in rendering tasks and in solving inverse graphics problems using gradients.


\end{abstract}


\begin{figure*}[!htb]
    \centering
    \begin{subfigure}[b]{0.3\textwidth}
        \centering
        \includegraphics[width=\textwidth]{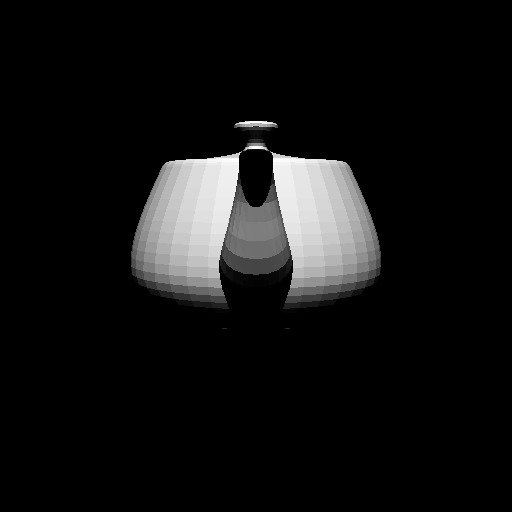}
        \caption*{}
    \end{subfigure}
    \hfill
    \begin{subfigure}[b]{0.3\textwidth}
        \centering
        \includegraphics[width=\textwidth]{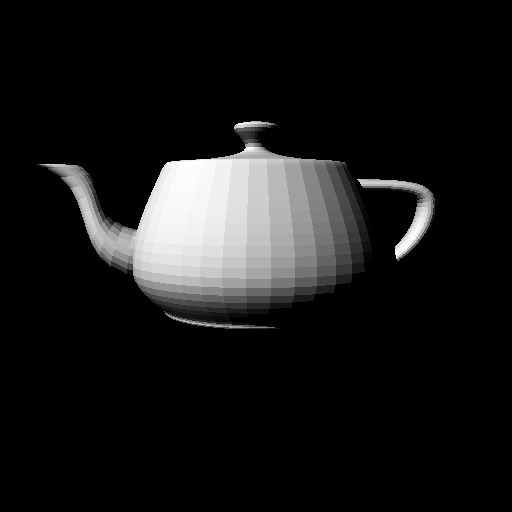}
        \caption*{}
    \end{subfigure}
    \hfill
    \begin{subfigure}[b]{0.3\textwidth}
        \centering
        \includegraphics[width=\textwidth]{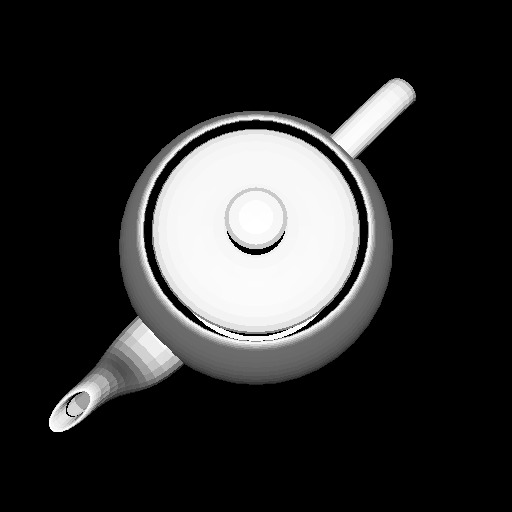}
        \caption*{}
    \end{subfigure}
    \caption{Utah Teapot Render from three different views. The camera definition shown in Listing~\protect\ref{lst:example_render} can be easily modified to generate all these views.}
    \label{fig:teapot_render}
\end{figure*}

\section{Introduction}
\label{intro}

Rendering is a technique of generating photo-realistic or non-photo-realistic 2D projections from 3D objects. As such, there are several algorithms for rendering complex scenes. One of the most popular techniques for photo-realistic rendering is ray tracing~\cite{Appel:1968:TSM:1468075.1468082}. For real-time rendering we use techniques like rasterization~\cite{Pineda:1988:PAP:54852.378457} are used.

Ray Tracing is a technique in computer graphics for rendering 3D graphics with complex light interactions. In this technique, rays are traced backward from the eye/camera to the light source(s). The ray can undergo reflection and refraction due to interactions with the objects in its path. This technique, however, is very computationally expensive and hence difficult to do in real-time. 

Since ray tracing leverages the properties of the materials of the objects in the scene, a natural extension to the rendering problem would be to extract the exact properties of the materials, lighting, and so on, given an image of a scene. This task is known as inverse rendering. Calculating analytic gradients for every single parameter of the scene is a very tedious process and prone to errors. This has made it a difficult task to present a general gradient-based inverse rendering method. As such, there is only one framework in our knowledge, redner~\cite{Li:2018:DMC}, which has been able to do so by using analytic gradients. However, we bypass this problem by using AD. There have also been attempts at making rasterization differentiable~\cite{softraster}; however, this involves making changes in the core technique which is against our design principles.

Rendering is a computationally expensive technique, and so it is generally done in static languages like C++. Developing software in such languages are incredibly time-consuming. Also, most languages lack the support of the state of the art automatic differentiation tools like Zygote~\cite{DBLP:journals/corr/abs-1810-07951}, Jax~\cite{jax}, which are generally implemented for high-level languages like Julia and Python. As such, it is challenging to develop differentiable renderers in those languages and then interface with popular deep learning software. Most of the other existing AD softwares, which the authors are familiar with, like CasADi~\cite{casadi} does not seamlessly integrate with packages, which means one needs to rewrite the software to use specific AD tools.

In this paper, we explore the idea of differentiability through a renderer, by leveraging the AD in Julia~\cite{bezanson2017julia}. We present a fully general renderer capable of handling complex scenes and able to differentiate through them. We do not rely on analytic gradients but use source-to-source AD to generate efficient gradient code in the backward pass. Our renderer contains very little code for integration with Zygote, and hence, in theory, we can plug in any other AD software written in Julia.

\section{Differentiable Ray Tracing}

There are several photo-realistic renderers available which contain a vast amount of implicit knowledge. Differentiation allows such renderers to make use of gradients to learn the inverse mapping from an image to its parameter space. However, as usual, it is challenging to compute efficient derivatives from a production-ready renderer, typically written in a performance language like C++. This provides the primary motivation for the development of RayTracer.jl. We develop an entire general-purpose ray tracer in a high-level numerical computation language. The presence of strong automatic differentiation libraries like Zygote.jl makes it trivial to compute efficient derivatives from the renderer. We present the performance gains we get on using Zygote as compared to Central Differencing in Section~\ref{sec:finite_diff}.

RayTracer.jl~\cite{RayTracer.jl} is a package for Differentiable Ray Tracing written to solve this particular issue. It relies heavily on the source-to-source automatic differentiation package, Zygote, for computing gradients with respect to arbitrary scene parameters. This package allows the user to configure the location of objects, lights, and a camera in the scene. This scene is then interpreted by the renderer to generate the image. RayTracer.jl is naturally interfaced with the deep learning library Flux \cite{Flux.jl-2018}, due to the common AD backend, for use in more complex differentiable pipelines.

Ray Tracing is primarily a non-differentiable operation. As such, any technique that is used to compute gradients for scene parameters by backpropagating through a ray tracer would be some approximation of gradients. In order to make our rendering differentiable, we sample only a single ray for every pixel on the screen. For better image quality, we should be sampling multiple rays for a single pixel value, but this would make differentiation a bit tricky. Since we have only one ray per pixel, it is bound to intersect with only a single triangle in the 3D scene. The color of a pixel is a weighted sum of colors of the intersection points. The computation of point color is a differentiable operation as it is either a plain color or a texture, calculated using the barycentric coordinates. Also, checking the intersection between a ray and a triangle involves solving a quadratic equation which is again differentiable. Since every operation is differentiable, we can easily backpropagate the errors through this. However, in case the ray intersects at a point close to the intersection of two triangles, the gradients are not correct. This is because these points in space are non-differentiable. We discuss this shortcoming in Section~\ref{sec:limitation}.

\section{Scene Rendering}

We first create a general-purpose renderer and then make use of efficient AD tools to make it completely differentiable. Hence, at its core, RayTracer is a fully-featured renderer. It contains functionalities for both raytracing and rasterization. Unlike most prior work in differentiable rendering, we do not make performance compromises in the forward pass (rendering) to allow gradient computation.

RayTracer gives much control to the user over the scene they want to render. The user controls the lighting in the scene, the shape, and materials of the objects and the camera configuration.

\noindent
\begin{minipage}{\linewidth}
\begin{lstlisting}[caption = {Rendering the Utah Teapot Model},
                   label = {lst:example_render},
                   captionpos = b,
                   language = Julia]
# Screen Size
screen_size = (w = 512, h = 512)
    
# Camera Setup
cam = Camera(
    lookfrom = Vec3(1.0f0, 10.0f0, -1.0f0),
    lookat = Vec3(0.0f0),
    vup = Vec3(0.0f0, 1.0f0, 0.0f0),
    vfov = 45.0f0,
    focus = 1.0f0,
    width = screen_size.w,
    height = screen_size.h
)
                 
origin, direction = get_primary_rays(cam)
    
# Scene
scene = load_obj("teapot.obj")
    
# Light Position
light = DistantLight(
    color = Vec3(1.0f0),
    intensity = 100.0f0,
    direction = Vec3(0.0f0, 1.0f0, 0.0f0)
)
                         
# Render the image
color = raytrace(
    origin,
    direction,
    scene,
    light,
    origin,
    2
)
\end{lstlisting}
\end{minipage}

\begin{figure*}[!htb]
    \centering
    \begin{subfigure}[c]{0.47\textwidth}
        \includegraphics[width=\textwidth]{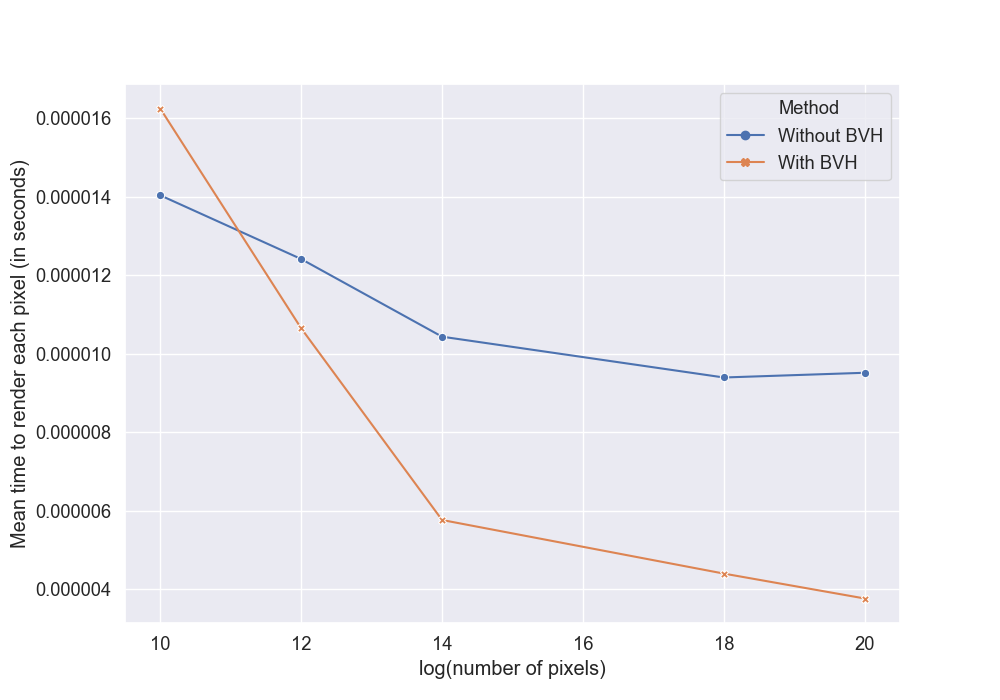}
        \caption{Performance Benchmarks}
        \label{fig:bvh_perf}
    \end{subfigure}
    \hfill
    \begin{subfigure}[c]{0.47\textwidth}
        \includegraphics[width=\textwidth]{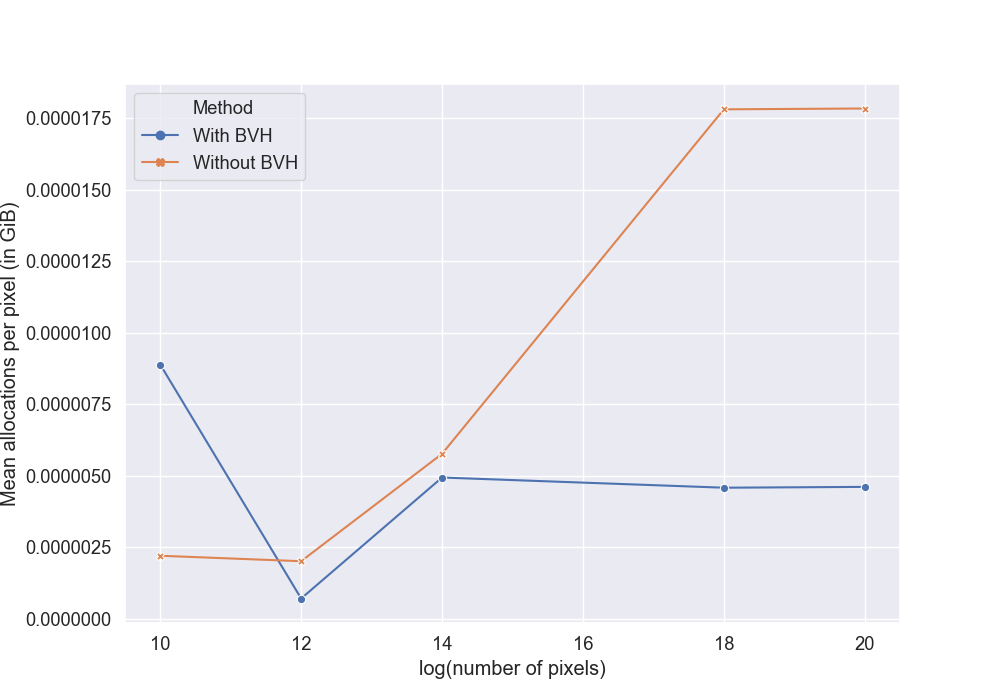}
        \caption{Memory Allocation Benchmarks}
        \label{fig:bvh_mem}
    \end{subfigure}
    \caption{Comparison between scenes rendered with and without BVH}
    \label{fig:bvh}
\end{figure*}

In this part, we will demonstrate the general pipeline for defining a 3D scene using RayTracer.jl and then rendering it. We are going to render the popular Utah Teapot model. We need to specify the 3D model in the form of a Vector of Objects. We can do it manually for custom scenes, or we could load it from a wavefront object (obj) file (MeshIO.jl provides support for additional file formats). Apart from the scene vector, we need to specify the camera configuration and the configuration of light(s). We summarize the entire code to render the teapot in Listing~\ref{lst:example_render}.

\section{Inverse Rendering}

The rendering problem is to project 3D scene parameters to form an image in the 2D plane. Inverse Rendering problem is just the opposite: generating a mapping from the 2D image back to the parameters of the 3D scene.

RayTracer.jl can be used to solve a variety of inverse graphics problems. Since the renderer can be used to compute the gradient with respect to any arbitrary parameter (as long as it is differentiable), we can then use any gradient-based optimization technique to optimize on that parameter. This allows us to propose a generalized Algorithm~\ref{alg:inv_render} which is capable of optimizing any differentiable parameter.

\begin{figure*}
    \centering
    \begin{subfigure}[c]{0.48\textwidth}
        \includegraphics[width=\linewidth, height=150px]{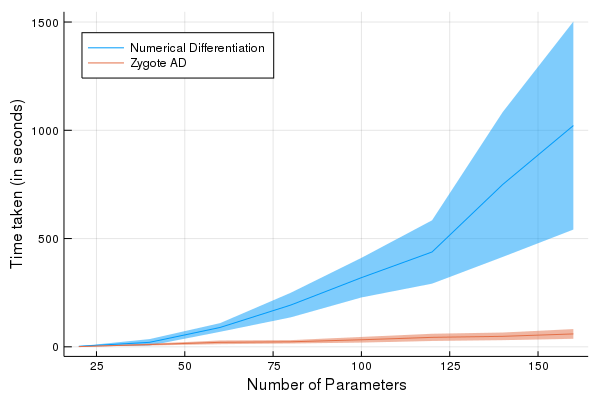}
        \caption{Time taken for the Backward Pass}
        \label{fig:ad_perf}
    \end{subfigure}
    \hfill
    \begin{subfigure}[c]{0.48\textwidth}
        \includegraphics[width=\linewidth, height=150px]{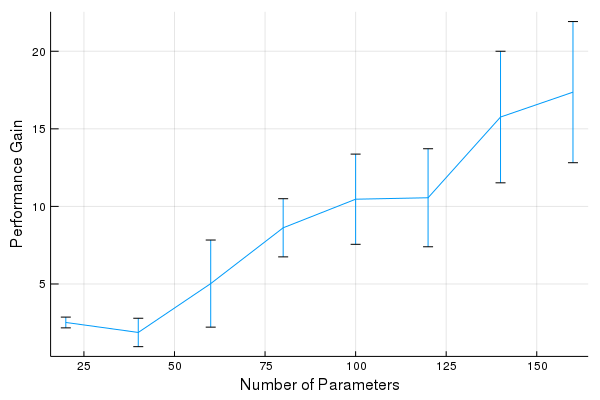}
        \caption{SpeedUp when using AD vs Numerical Differentiation}
        \label{fig:ad_perf_err}
    \end{subfigure}
    \caption{Comparison between Automatic Differentiation and Numerical Differentiation}
    \label{fig:ad_better}
\end{figure*}

\begin{algorithm}[!htb]
\DontPrintSemicolon
\SetAlgoLined
\KwIn{Initial Guess of the Scene Parameters, Maximum Number of Iterations, Tolerance in Loss}
\KwOut{Optimized set of Scene Parameters}
params $\gets$ Initial Guess of Scene Parameters\;
tolerance $\gets$ Tolerance in Loss\;
max\_iter $\gets$ Maximum Number of Iterations\;
converged $\gets$ false\;
iter $\gets$ 0\;
\While{not \textbf{converged} or iter < max\_iter}{
    loss $\gets$ mean\_squared\_loss (render\_image (params), target\_img)\;
    gs $\gets$ gradient(loss)\;
    \For{param in params}{
        update! (optimizer, param, gs[param])\;
    }
    \If{loss < tolerance}{
        converged $\gets$ true\;
    }
    iter $\gets$ iter + 1\;
}
\Return{params}\;
\caption{Gradient Based Optimization of Scene Parameters}
\label{alg:inv_render}
\end{algorithm}

\section{Experiments}

In this section we showcase our differentiable renderer in some benchmarking and toy inverse rendering problems. Using the following experiments we demonstrate the use of gradients obtained via AD to recover the camera, material and lighting parameters for a scene. In the inverse rendering experiments we make use of the Adam optimizer as described in \cite{kingma2014adam}. We interface the raytracer with Flux to use these optimizers. As an alternative, we have also tested the functioning of our package with the optimizers present in Optim\footnote{\raggedright{We provide an example at \href{https://zenodo.org/record/1442781}{examples/optim\_compatibility.jl}}} \cite{K2018Optim}.

\subsection{Accelerating the Rendering using Acceleration Structures}

To accelerate the rendering process, we support an acceleration structure, Bounding Volume Hierarchy (BVH)\cite{Kay:1986:RTC:15922.15916}. We follow the same API for ray tracing using these accelerators. In this case, instead of passing a Vector of Objects, we wrap it in a BoundingVolumeHierarchy object and pass it. So in order to use this, we would have to change the scene variable in Listing~\ref{lst:example_render} to \lstinline{BoundingVolumeHierarchy(...)}.

In this section, we provide a comparison between the performance gains and memory allocation benefits of using BVH. We use a mesh of 137 triangles centered at the origin as the scene. We increase the screen size and hence increasing the total number of pixels (and therefore primary rays) in the scene\footnote{Even though it might seem that increasing the number of objects in the scene would be a better metric for comparison, it is immensely difficult to make that comparison. This is primarily because the same scene in a different configuration will have a different render time}.

We present the benefits of using BVH in Figure~\ref{fig:bvh}. We are able to get to reduce the total allocations (Figure~\ref{fig:bvh_mem}) and also get a significant performance boost\footnote{\raggedright{We provide the code for reproducing the experiment in \href{https://zenodo.org/record/1442781}{examples/performance\_benchmarks.jl}}} (Figure~\ref{fig:bvh_perf}). Note that we only get exponential benefits in terms of memory and performance when the number of pixels is of reasonable size, for example in case of $128 \times 128$ or better resolution images. For smaller images, using BVH might end up slowing down the rendering process.

\subsection{Comparison with Finite Differencing}
\label{sec:finite_diff}

In this section, we demonstrate the performance gain of using source-to-source AD backend versus Central Finite Differencing. We provide a convenience function for finite differencing for gradient testing purposes. The $\delta$ for calculating derivatives using the equation $\frac{df}{dx} = \frac{f(x + \delta) - f(x - \delta)}{2\delta}$ is fixed by comparing these values with the values obtained from Zygote. Through this experiment, we show that our method is exponentially better than numerical differentiation.

For comparing the two differentiation techniques, we run five independent trials and present the mean runtimes (with the standard deviation) in Figure~\ref{fig:ad_better}. We fix the position of the camera and light and randomly generate the triangles present in the scene. Every new triangle added to the scene, adds 20 additional parameters with respect to which we must compute the derivatives. We use the mean squared loss function to compute the scalar loss and backpropagate.

Figure~\ref{fig:ad_perf_err} shows that we get significant speedups for a reasonable number of parameters. The nature of speedup shown suggests that it is nearly infeasible to use numerical differentiation when the number of parameters exceeds 50. In most practical applications of differentiable mesh rendering involving neural networks, we will have at least thousands of parameters. In such cases, our AD-based solution will be able to compute the derivatives in a reasonable time.

\subsection{Calibration of Camera Parameters}
\label{sec:calcam}

\begin{figure*}[!htb]
    \centering
    \begin{subfigure}[c]{0.45\textwidth}
        \includegraphics[width=\linewidth]{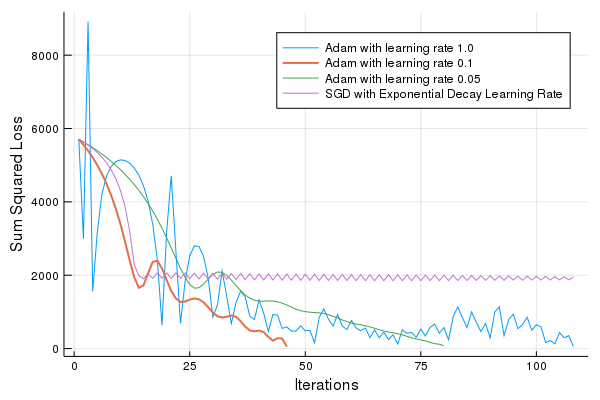}
        \caption{Loss Values over the Light Configuration Optimization Process}
        \label{fig:cam_curve}
    \end{subfigure}
    \hfill
    \begin{subfigure}[c]{0.45\textwidth}
        \begin{subfigure}[c]{0.45\textwidth}
            \centering
            \includegraphics[width=\textwidth]{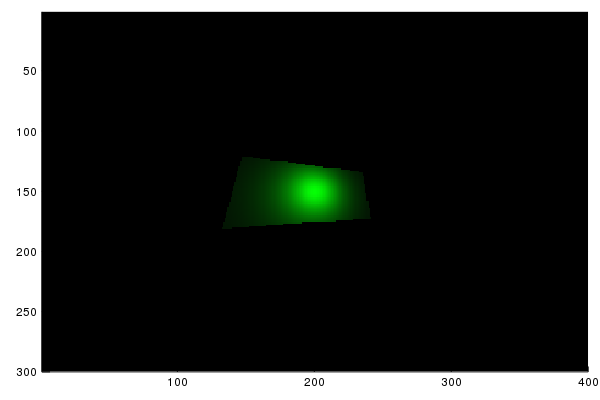}
            \caption{Image with uncalibrated camera}
            \label{fig:cam_guess}
        \end{subfigure}
        \hfill
        \begin{subfigure}[c]{0.45\textwidth}
            \centering          
            \includegraphics[width=\textwidth]{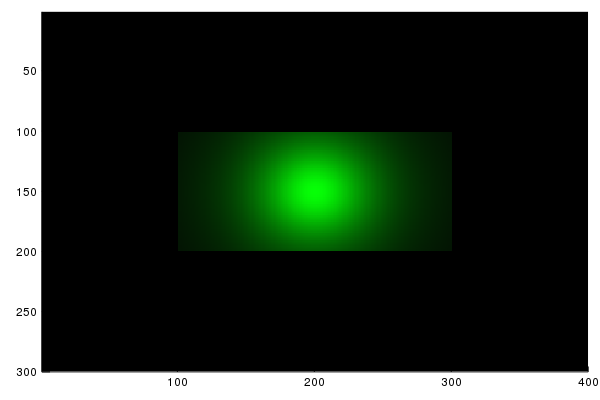}
            \caption{The image to be reconstructed}
            \label{fig:cam_target}
        \end{subfigure}
        \centering
        \hfill
        \begin{subfigure}[c]{0.45\textwidth}
            \centering
            \includegraphics[width=\textwidth]{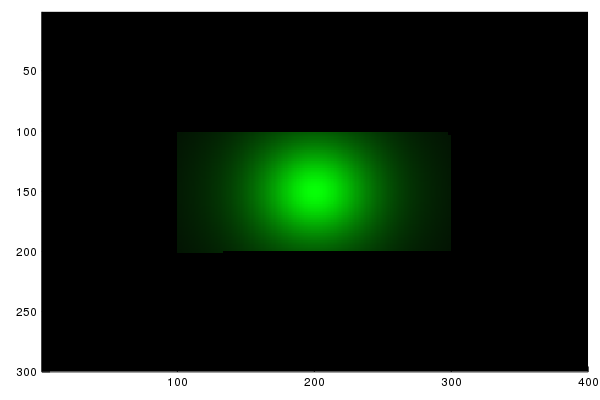}
            \caption{Image obtained after the optimization}
        \end{subfigure}
        \hfill
        \begin{subfigure}[c]{0.45\textwidth}
            \centering
            \includegraphics[width=\textwidth]{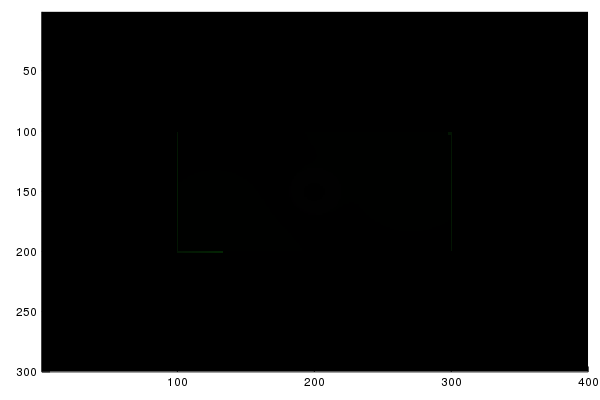}
            \caption{Difference between reconstructed and target image}
        \end{subfigure}
    \end{subfigure}
    \caption{Calibration of Camera Parameters to reconstruct Image~\ref{fig:cam_target} from Image~\ref{fig:cam_guess}}
    \label{fig:cam_invrender}
\end{figure*}

In this experiment we start with the image of a rectangle (Listing~\ref{lst:light_config}) under some configuration of the Camera model (Listing~\ref{lst:cam_opt}). Since RayTracer supports only two primitive shapes - Spheres and Triangles, we need to triangulate the rectangle.

\begin{lstlisting}[caption = {Configuration of the Scene for Experiment~\ref{sec:calcam}},
                   label = {lst:light_config},
                   captionpos = b,
                   language = Julia]
scene = [
    Triangle(
        Vec3( 20.0,  10.0, 0.0),
        Vec3( 20.0, -10.0, 0.0),
        Vec3(-20.0,  10.0, 0.0),
        Material(color_diffuse = Vec3(0.0, 1.0, 0.0))),
    Triangle(
        Vec3(-20.0, -10.0, 0.0),
        Vec3( 20.0, -10.0, 0.0),
        Vec3(-20.0,  10.0, 0.0),
        Material(color_diffuse = Vec3(0.0, 1.0, 0.0)))
]

light = PointLight(
    Vec3(1.0, 0.0, 0.0),
    100000.0,
    Vec3(0.0, 0.0, -10.0)
)
\end{lstlisting}

\noindent
\begin{minipage}{\linewidth}
\begin{lstlisting}[caption = {Camera Parameters to be Reconstructed},
                   label = {lst:cam_opt},
                   captionpos = b,
                   language = Julia]
camera_target =
    Camera(
        Vec3(0.0, 0.0, -30.0),
        Vec3(0.0, 0.0,   0.0),
        Vec3(0.0, 1.0,   0.0),
        90.0,
        1.0,
        screen_size...
    )
\end{lstlisting}
\end{minipage}

\noindent
\begin{minipage}{\linewidth}
\begin{lstlisting}[caption = {Initial Guess of the Camera Parameters},
                   label = {lst:cam_initial},
                   captionpos = b,
                   language = Julia]
camera_guess =
    Camera(
        Vec3(5.0, -4.0, -20.0),
        Vec3(0.0,  0.0,   0.0),
        Vec3(0.0,  1.0,   0.0),
        90.0,
        3.0,
        screen_size...
    )
\end{lstlisting}
\end{minipage}

We aim to reconstruct the image of this rectangle (Figure~\ref{fig:cam_target}) by modifying the focus and the location of the camera. We assume that all the other parameters of the model, like the light configuration (Listing~\ref{lst:light_config}) and the position of objects are known apriori. We use algorithm~\ref{alg:inv_render} for optimizing the parameters. We make an initial guess of the parameters and initialize the camera (Listing~\ref{lst:cam_initial}).

As our loss function, we use the mean squared difference between rendered and target images, each with $300\times400$ pixels having fractional RGB values. We minimize loss with the Adam optimizer, with learning rate 0.1, and declare the optimization to have converged if loss falls below $100$ (where the initial loss is $5705.98$). Figure~\ref{fig:cam_invrender} shows the optimization steps. We present the various learning rates and optimizers we experimented with in Figure~\ref{fig:cam_curve}.

\subsection{Optimizing the Light Source}
\label{sec:light_source}

\begin{figure*}[!htb]
    \centering
    \begin{subfigure}[c]{0.45\textwidth}
        \includegraphics[width=\linewidth]{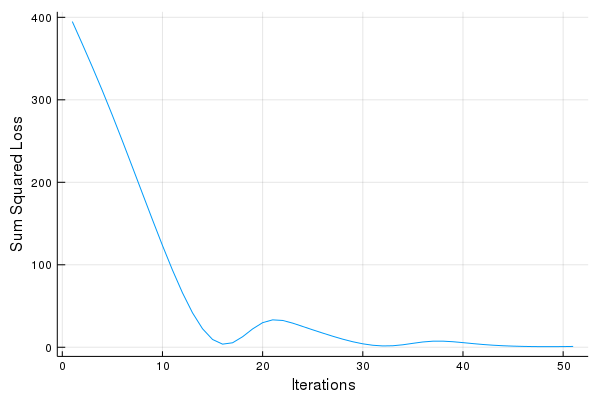}
        \caption{Loss Values over the Light Configuration Optimization Process}
        \label{fig:loss_plot_light}
    \end{subfigure}
    \hfill
    \begin{subfigure}[c]{0.45\textwidth}
        \begin{subfigure}[c]{0.45\textwidth}
            \centering
            \includegraphics[width=\textwidth]{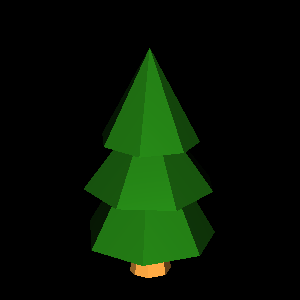}
            \caption{The image to be reconstructed}
            \label{fig:target_light}
        \end{subfigure}
        \hfill
        \begin{subfigure}[c]{0.45\textwidth}
            \centering
            \includegraphics[width=\textwidth]{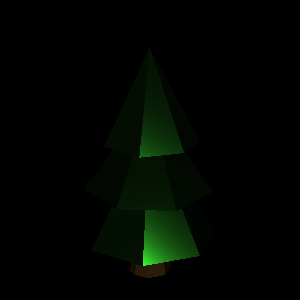}
            \caption{Initial Guess of Lighting Parameters}
            \label{fig:guess_light}
        \end{subfigure}
        \centering
        \hfill
        \begin{subfigure}[c]{0.45\textwidth}
            \centering
            \includegraphics[width=\textwidth]{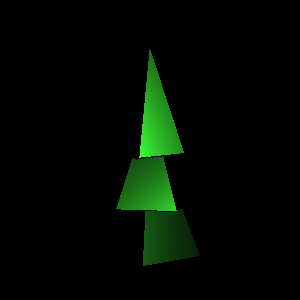}
            \caption{Image produced after 10 iterations}
        \end{subfigure}
        \hfill
        \begin{subfigure}[c]{0.45\textwidth}
            \centering
            \includegraphics[width=\textwidth]{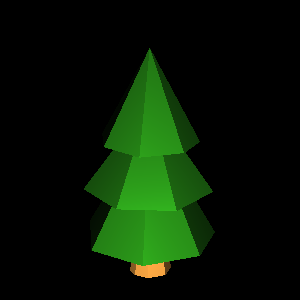}
            \caption{Image with converged parameters}
        \end{subfigure}
    \end{subfigure}
    \caption{Optimization of the lighting conditions to reconstruct Image~\ref{fig:target_light} from Image~\ref{fig:guess_light}}
    \label{fig:light_position}
\end{figure*}

In this experiment, we describe our solution to the inverse lighting problem. This problem involved predicting the configuration of the light source(s) in the scene given a target image (Figure~\ref{fig:target_light}). We know the exact geometry and surface properties of the objects, as well as the camera configuration. Listing~\ref{lst:initial_config_expt2} describes the known configurations.

\noindent
\begin{minipage}{\linewidth}
\begin{lstlisting}[caption = {Configuration of the Scene for Experiment~\ref{sec:light_source}},
                   label = {lst:initial_config_expt2},
                   captionpos = b,
                   language = Julia]

screen_size = (w = 128, h = 128)

camera = Camera(
    Vec3(0.0f0, 6.0f0, -10.0f0),
    Vec3(0.0f0, 2.0f0,  0.0f0),
    Vec3(0.0f0, 1.0f0,  0.0f0),
    45.0f0,
    0.5f0,
    screen_size...
)

scene = load_obj("tree.obj")
\end{lstlisting}
\end{minipage}

The object in our scene is a tree. We start with arbitrary lighting (Listing~\ref{lst:lgt_guess}) condition and then iteratively improve the lighting using Algorithm~\ref{alg:inv_render}. We present the loss curve and the images generated during the optimization process in Figure~\ref{fig:light_position}.

\noindent
\begin{minipage}{\linewidth}
\begin{lstlisting}[caption = {Target Lighting Conditions},
                   label = {lst:lgt_target},
                   captionpos = b,
                   language = Julia]
light_target = PointLight(
    Vec3(1.0f0, 1.0f0, 1.0f0),
    20000.0f0,
    Vec3(1.0f0, 10.0f0, -50.0f0)
)
\end{lstlisting}
\end{minipage}

\noindent
\begin{minipage}{\linewidth}
\begin{lstlisting}[caption = {Initial Guess of Lighting Conditions},
                   label = {lst:lgt_guess},
                   captionpos = b,
                   language = Julia]
light_guess = PointLight(
    Vec3(1.0f0, 1.0f0, 1.0f0),
    1.0f0,
    Vec3(-1.0f0, -10.0f0, -50.0f0)
)
\end{lstlisting}
\end{minipage}

\subsection{Retrieving Color of Materials}
\label{sec:mat_color}

\begin{figure}[!htb]
    \centering
    \begin{subfigure}[c]{0.23\textwidth}
        \centering
        \includegraphics[width=\textwidth]{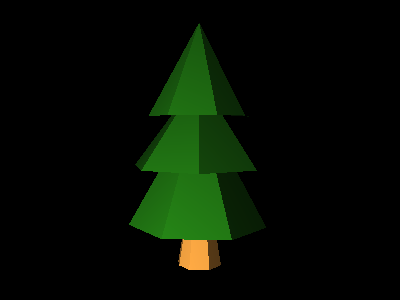}
        \caption{Image to be reconstructed}
        \label{fig:target_material}
    \end{subfigure}
    \hfill
    \begin{subfigure}[c]{0.23\textwidth}
        \centering
        \includegraphics[width=\textwidth]{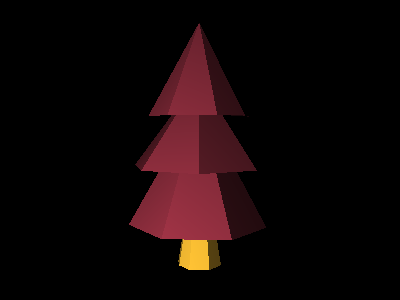}
        \caption{Initial Guess of the Materials}
        \label{fig:guess_material}
    \end{subfigure}
    \centering
    \hfill
    \begin{subfigure}[c]{0.23\textwidth}
        \centering
        \includegraphics[width=\textwidth]{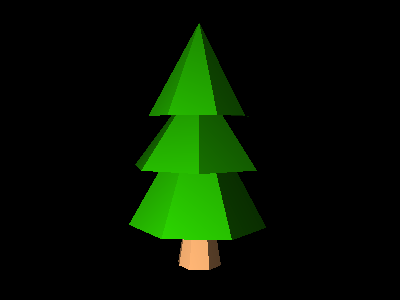}
        \caption{Image produced after just 1 iteration}
        \label{fig:single_step}
    \end{subfigure}
    \hfill
    \begin{subfigure}[c]{0.23\textwidth}
        \centering
        \includegraphics[width=\textwidth]{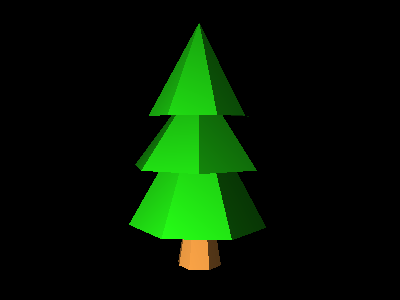}
        \caption{Image obtained after 35 iterations}
    \end{subfigure}
    \caption{Optimization of the materials of the mesh to reconstruct Image~\ref{fig:target_material} from Image~\ref{fig:guess_material}}
    \label{fig:mat_correction}
\end{figure}

RayTracer.jl can also be used to recover the properties of the material of a mesh. For this experiment, we shall use the same tree mesh from Experiment~\ref{sec:light_source}. We are going to optimize the diffuse color of the mesh. We use the position of the camera and the lighting conditions, as mentioned in Listing~\ref{lst:config_mat}.

The significant difference of this optimization from the prior experiments is that the color can only take values between $0.0$ and $1.0$. After every iteration, we clamp the value of the diffuse color. Hence, we update our parameters using projected gradient descent.

We use the sum squared loss function. We minimize the loss with Adam optimizer, with a learning rate of 0.05. The convergence of the model is quite fast, and we get a good approximation of the parameters just after a single optimizer step (Figure~\ref{fig:single_step}).

\noindent
\begin{minipage}{\linewidth}
\begin{lstlisting}[caption = {Configuration of the Scene for Experiment~\ref{sec:mat_color}},
                   label = {lst:config_mat},
                   captionpos = b,
                   language = Julia]
screen_size = (w = 400, h = 300)

light = PointLight(
    Vec3(1.0f0),
    1000000.0f0,
    Vec3(0.15f0, 0.5f0, -10.5f0)
)

cam = Camera(
    Vec3(-2.0f0, 2.0f0, -5.0f0),
    Vec3( 0.0f0, 1.7f0,  0.0f0),
    Vec3( 0.0f0, 1.0f0,  0.0f0),
    45.0f0,
    1.0f0,
    screen_size...
)

scene = load_obj("tree.obj")
\end{lstlisting}
\end{minipage}

\section{Current Limitations}
\label{sec:limitation}

Despite the success of our approach in solving a variety of inverse graphics problems, we fail to deal with non-differentiable problems. One such instance would be estimating the proper geometry of an object given an image. Such problems are non-differentiable due to a large number of discrete choices in the position of the triangle vertices. Hence, trying to optimize such parameters generally causes them to diverge. The other cases which we cannot handle properly are secondary lighting (shadows) and global illumination. Most of these cases can be dealt with, similar to the way proposed by \cite{Li:2018:DMC}, but that leads to a significant slowdown to the rendering of the scene.


\section{Conclusion}

In conclusion, we have shown how Julia can be leveraged to build differentiable systems. We have presented which to the best of our knowledge is the first differentiable renderer which uses source-to-source AD. We have used a set of toy examples to demonstrate the ability of our renderer to reconstruct scenes (which are differentiable) from only a single image. This also shows that this renderer can be used in differentiable programming pipelines which involve image generation.

\bibliographystyle{IEEEtran}
\bibliography{references}


\end{document}